\documentclass[twocolumn]{aa} 

\usepackage{graphicx}
\usepackage{txfonts}
\usepackage{natbib}
\begin{document}
   \title{W49A: A starburst triggered by expanding shells}


   \author{T.-C. Peng\inst{1}
          \and F. Wyrowski\inst{1}
          \and F. F. S. van der Tak\inst{2}
          \and K. M. Menten\inst{1}
          \and C. M. Walmsley\inst{3}
          }

   \institute{Max-Planck-Institut f\"ur Radioastronomie (MPIfR),
              Auf dem H\"ugel 69, 53121 Bonn, Germany \\
              \email{tcpeng@mpifr-bonn.mpg.de}
              \and SRON Netherlands Institute for Space Research, 9747 AD Groningen, Netherlands
              \and INAF Osservatorio Astrofisico di Arcetri, I-50125 Firenze, Italy \\
             }

   \date{}

 
  \abstract
   {}
   {W49A is a giant molecular cloud which harbors some of the most luminous embedded clusters in the Galaxy. However, the explanation for this starburst-like phenomenon is still under debate.}
   {We investigated large-scale Spitzer mid-infrared images together with a Galatic Ring Survey $^{13}$CO $J=1-0$ image, complemented with higher resolution ($\sim 11\arcsec$) $^{13}$CO $J=2-1$ and C$^{18}$O $J=2-1$ images over a $\sim15\times13$ pc$^2$ field obtained with the IRAM 30m telescope.}
   {Two expanding shells have been identified in the mid-infrared images, and confirmed in the position-velocity diagrams made from the $^{13}$CO $J=2-1$ and C$^{18}$O $J=2-1$ data. The mass of the averaged expanding shell, which has an inner radius of $\approx3.3$ pc and a thickness of $\approx 0.41$ pc, is about $1.9\times10^{4}$ M$_{\sun}$. The total kinetic energy of the expanding shells is estimated to be $\sim10^{49}$ erg which is probably provided by a few massive stars, whose radiation pressure and/or strong stellar winds drive the shells. The expanding shells are likely to have a common origin close to the two ultracompact H{\sc ii} regions (source O and source N), and their expansion speed is estimated to be $\sim5$ km s$^{-1}$, resulting in an age of $\sim3$--$7\times10^{5}$ years. In addition, on larger ($\sim35\times50$ pc$^2$) scales, remnants of two gas ejections have been identified in the $^{13}$CO $J=1-0$ data. Both ejections seem to have the same center as the expanding shells with a total energy of a few times $10^{50}$ erg. The main driving mechanism for the gas ejections is unclear, but likely related to the mechanism which triggers the starburst in W49A.}
   {}


   \maketitle
%

\section{Introduction}

A variety of processes have been suggested that might trigger the formation of stars, namely large-scale supernova explosions or smaller-scale stellar winds, outflow shocks, or ionization fronts \citep[see][]{McKee2007,Elmegreen1998}. Possible triggers are newly formed massive stars, which provide energetic feedback to the interstellar medium (ISM) in the form of expanding H{\sc ii} regions or strong stellar winds. A recent study of the ubiquitous dust bubbles observed in the mid-infrared (MIR, 5--40 $\mu$m) in the Galactic Legacy Infrared Mid-Plane Survey Extraordinaire \citep[GLIMPSE;][]{Benjamin2003} shows that, given the positional coincidence with known H{\sc ii} regions, many MIR bubbles are indeed produced by hot young stars, and are likely driven by expanding H{\sc ii} shells \citep{Churchwell2006,Watson2008}. In addition, for reasonable assumptions on the ambient density, the numerical simulations of expanding H{\sc ii} regions carried out by \citet{Hosokawa2005,Hosokawa2006} agree with some observations, e.g., \citet{Watson2008}.

   \begin{figure*}
   \centering
   \includegraphics[angle=-90,width=\textwidth]{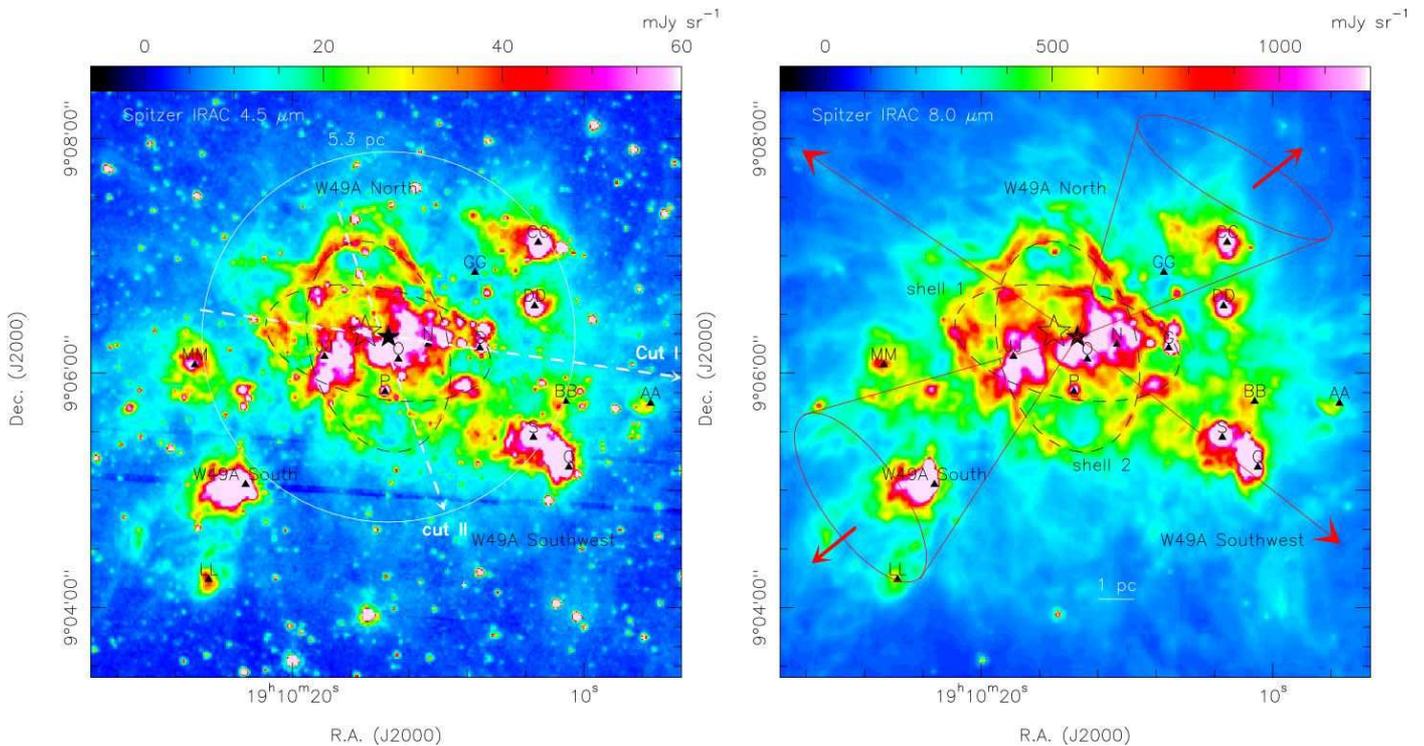}
   \caption{The Spitzer-GLIMPSE IRAC 4.5 $\mu$m and 8.0 $\mu$m images of W49A. Black triangles denote the UC H{\sc ii} regions observed by \citet{DePree1997}. The black dashed ellipses are the suggested expanding shells (shells 1 and 2), and the black filled star represents the center of the expanding shells and gas ejections (see text). The larger black unfilled star represents the embedded cluster center according to \citet{Alves2003}. The white circle in the left image indicates the maximum radius (5.3 pc) used in the circular PV diagram shown in Figure \ref{Fig2}, and the two white dashed lines represent the linear PV cuts (Fig. \ref{Fig2}). Red lines and arrows in the right image indicate the directions of gas ejections.}
              \label{Fig1}
    \end{figure*}

The giant molecular cloud (GMC) associated with the powerful radio continuum source W49A is part of the W49 complex which also includes the supernova remnant W49B. At a distance of 11.4 kpc \citep{Gwinn1992}, W49A has an infrared luminosity of $>10^{7}$ L$_{\sun}$ and a total mass of $\sim10^{6}$ M$_{\sun}$ \citep{Sievers1991}, and contains over a dozen of ultracompact H{\sc ii} (UC H{\sc ii}) regions \citep{DePree1997,DePree2000} as well as the brightest water maser cluster in our Galaxy \citep[$\sim$1 L$_{\sun}$ alone in the 22 GHz line,][]{Genzel1978}. The embedded massive stellar cluster found in W49A indicates a somewhat more evolved stage of massive star formation in this region \citep{Alves2003}. Many authors have tried to explain why W49A, a Galactic mini-starburst, has formed so many massive stars in such a short time. \citet{Welch1987} concluded the existence of a large-scale inside-out gravitational collapse based on their interpretation of the HCO$^+$ spectra measured toward the ring-like configuration of UC H{\sc ii} regions. However, \citet{Serabyn1993} argued that the two components of the double-peak line profile seen in CS and C$^{34}$S come from different clouds, and suggested that massive star formation in W49A is triggered by a large-scale cloud-cloud collision. Recently, it has been argued that one cannot distinguish between a global collapse model of a large cloud and a multiple-cloud model, given the existing HCO$^{+}$ and C$^{18}$O data \citep{Williams2004}. In this paper, we present new evidence of expanding shells in W49A delineated by dense clumps within these shells. The expanding shells provide a natural explanation for the observed molecular line profiles, and triggered massive star formation in W49A.

%

\section{Observations}

Observations of the $^{13}$CO $J=2-1$ and C$^{18}$O $J=2-1$ transitions toward W49A were carried out in 2006 November and 2007 November using the IRAM 30 m telescope on Pico Veleta in Spain. The focus was checked on Uranus, and the pointing was checked on the 
strong continuum emission from the BL Lac object 1749+096 and Uranus every hour. The accuracy of the pointing was $\sim 2\arcsec$--$3\arcsec$ with a telescope tracking accuracy of 1\arcsec. The chopper wheel method was used to calibrate the data.

We used the HEterodyne Receiver Array, HERA \citep{Schuster2004}, to obtain large On-The-Fly (OTF) maps of W49A. HERA is a nine-pixel
receiver with dual polarizations in a square center-filled 3$\times$3 array, and can be tuned to different sky frequencies with 
different backends attached. In particular, simultaneous observations of $^{13}$CO $J=2-1$ and C$^{18}$O $J=2-1$ are possible. A dewar rotation angle of 9.6\degr of HERA was chosen to have Nyquist-sampled maps. Maps sizes of $\sim$ 4\farcm5$\times$4\farcm0 were 
obtained for the $^{13}$CO and C$^{18}$O $J=2-1$ lines with an HPBW of $\sim$ 11\arcsec. Every scan in right ascension was followed by one in declination in order to reduce striping patterns. The calibration was done for each OTF subscan and reduced in the MIRA software of the GILDAS\footnote{http://www.iram.fr/IRAMFR/GILDAS} 
package. OFF scans on the reference position ($-$600\arcsec, $-$600\arcsec) were done before and after each OTF subscan. The Versatile SPectrometer Array (VESPA) was chosen as the backend for HERA configured to a bandwidth of 160 MHz for each receiver element and 0.32 km s$^{-1}$ resolution. The data were reduced using the standard procedures in the GILDAS package, and were corrected to main beam temperature $T_{\rm{MB}}$ units from the antenna temperatures $T_{\rm{A}}^{*}$ (i.e., $T_{\rm{MB}}$ = $T_{\rm{A}}^{*}$/$\eta_{\rm{MB}}$) which were only corrected for atmospheric absorption, rear spillover, and ohmic losses. The efficiency, $\eta_{\rm{MB}}$, is given by the ratio between the main beam efficiency (named $B_{\rm{eff}}$; it is 0.51 at 220 GHz in the IRAM documentation\footnote{http://www.iram.fr/IRAMES}), and the forward efficiency $F_{\rm{eff}}$ (0.91 at 220 GHz). 

\section{Results}

Figure \ref{Fig1} shows evidence of shell structures in W49A in the 4.5 $\mu$m and 8.0 $\mu$m images from the GLIMPSE archive. They were obtained with the Spitzer Infrared Array Camera \citep[IRAC;][]{Fazio2004}. The 4.5 $\mu$m band, which contains a pure rotational transition of H$_2$, mainly traces the distribution of warm molecular hydrogen. The 8.0 $\mu$m band, which includes two aromatic infrared features at 7.7 $\mu$m and 8.6 $\mu$m, is believed to be dominated by the emission of polycyclic aromatic hydrocarbon (PAH) molecules primarily excited by UV radiation. The angular resolution of IRAC ranges from $\sim 1\farcs5$ at 3.6 $\mu$m to $\sim 1\farcs9$ at 8.0 $\mu$m. In two other IRAC bands (3.6 $\mu$m and 5.8 $\mu$m), however, the same structures can also be clearly seen. The shells were first identified by eye and then investigated in the molecular data cube. Two MIR shells have been identified (Fig. \ref{Fig1}) with long axes of $55\arcsec$--$58\arcsec$ and short axes of $28\arcsec$--$35\arcsec$, which implies an average size of $\sim2$--$3$ pc in radius. The average thickness of the MIR shells is estimated to be $7\farcs5\pm1\farcs9$ ($0.4\pm0.1$ pc) from fits of Gaussian profiles to the shells. These two MIR shells lie in the east-west and north-south directions, and have a similar size and elliptical shape. In addition, they seem to share the same center.

   \begin{figure*}
   \centering
   \includegraphics[angle=-90,width=\textwidth]{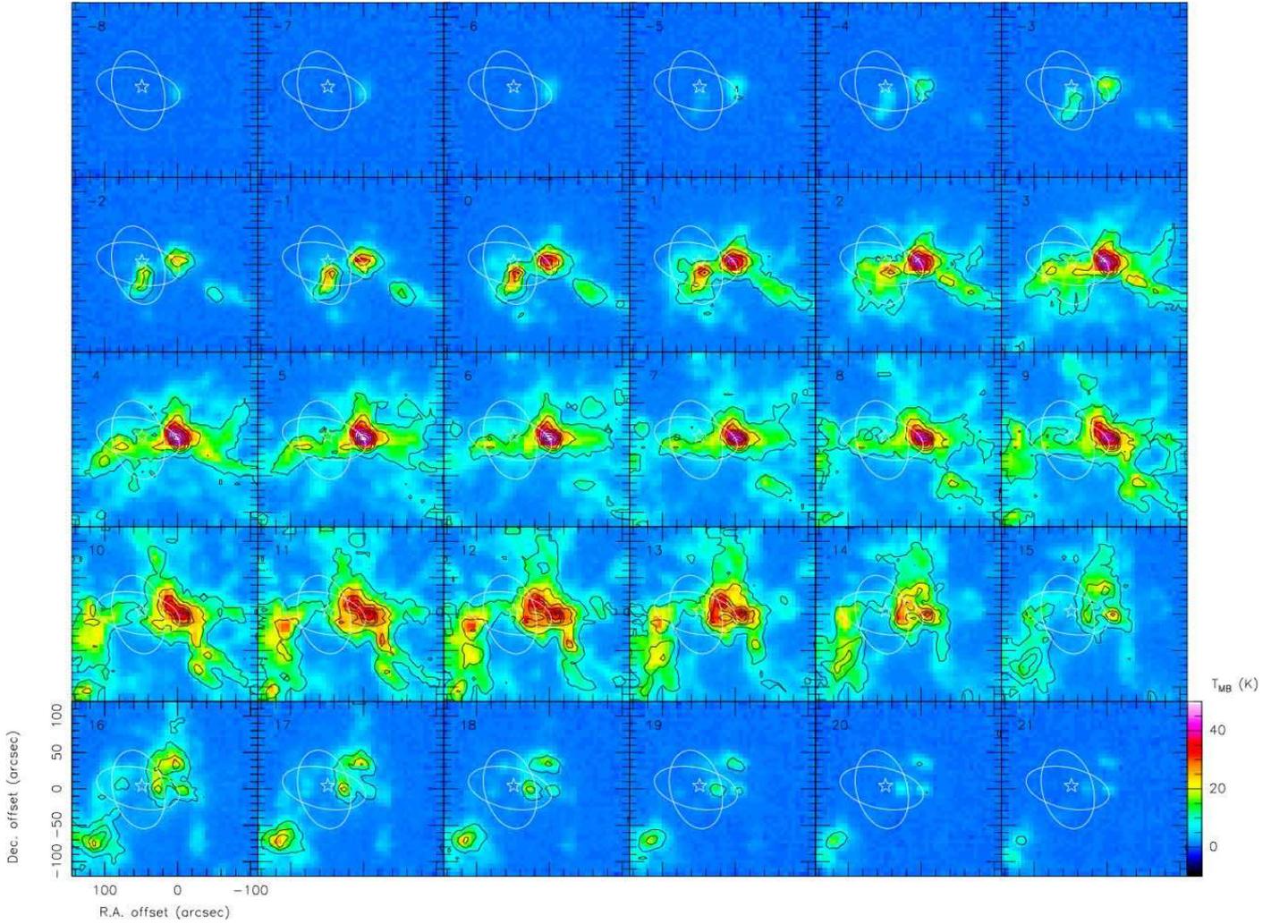}
   \caption{The IRAM 30m HERA channel maps of $^{13}$CO $J=2-1$ (images) overlaid with the C$^{18}$O $J=2-1$ emission in black contours with a resolution of 11\arcsec. The contours are plotted from 1 to 15 K in steps of 2 K. The white ellipses represent two suggested shells, and the white star marks the shell expansion center. The ($0\arcsec,0\arcsec$) position is located at the region around source G.}
              \label{Fig4}
    \end{figure*}

\begin{figure*}
   \centering
   \includegraphics[angle=-90,width=\textwidth]{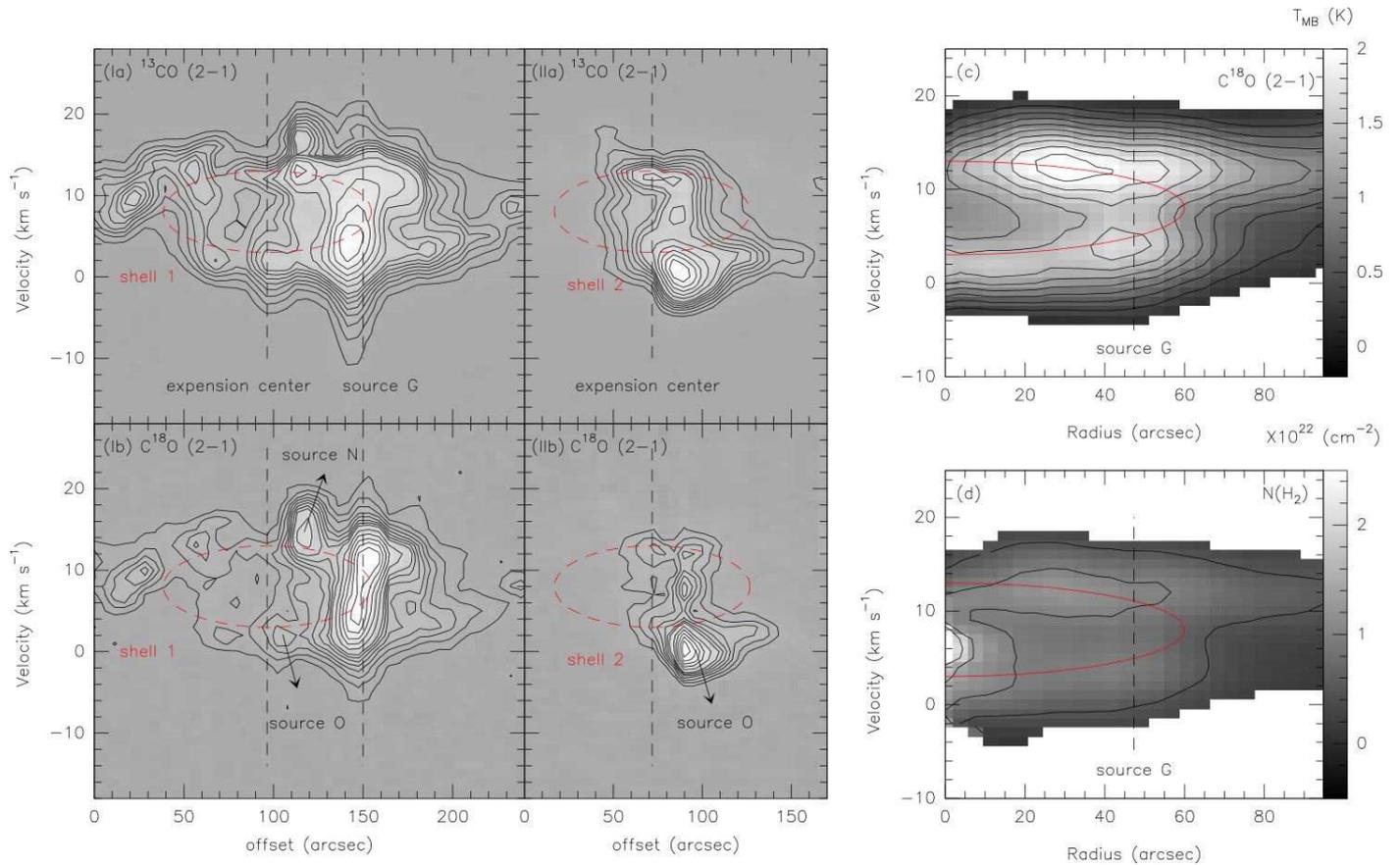}
   \caption{Left panel: The linear PV diagrams of $^{13}$CO $J=2-1$ and C$^{18}$O $J=2-1$ corresponding to Cut I (E$-$W) and Cut II (N$-$S) shown in Figure \ref{Fig1}. The suggested shell expansion center and source G are marked. The red dashed ellipses indicate the expansion speed of 5 km s$^{-1}$ and the radius corresponding to the two MIR shells marked in Figure \ref{Fig1}. The clumps associated with source O and source N are marked. (Ia)-(Ib) Black contours represent the main-beam brightness temperatures $T_{\rm MB}$ from $5\%$ (3 $\sigma$ for C$^{18}$O and 10 $\sigma$ for $^{13}$CO) to $35\%$ in steps of $5\%$ of the peak value, and the subsequent contours are running from $40\%$ to $90\%$ in steps of $10\%$ of the peak value. (IIa)-(IIb) Black contours represent $T_{\rm MB}$ from $10\%$ (3 $\sigma$ for C$^{18}$O and 10 $\sigma$ for $^{13}$CO) to $35\%$ in steps of $5\%$ of the peak value, and the subsequent contours are running from $40\%$ to $90\%$ in steps of $10\%$ of the peak value. (c) The circular PV diagram of C$^{18}$O $J=2-1$. The radius axis is the distance to the shell expansion center, and the temperature is averaged in a given radius. Black contours represent the main-beam brightness temperatures of C$^{18}$O $J=2-1$ from 0.4 to 2.0 K in steps of 0.2 K. (d) The H$_2$ column density of an averaged shell calculated in the LTE assumption. Black contours represent the H$_2$ column densities of $5.0\times10^{21}$ cm$^{-2}$, $1.0\times10^{22}$ cm$^{-2}$, $1.5\times10^{22}$ cm$^{-2}$, and $2.0\times10^{22}$ cm$^{-2}$. The shell with a radius size of 3.3 pc and an expansion speed of 5 km s$^{-1}$ is indicated in red ellipses.}
              \label{Fig2}
    \end{figure*}

However, it is difficult to identify any similar shell structure in the channel maps of the $^{13}$CO and C$^{18}$O $J=2-1$ lines (see Fig. \ref{Fig4}) because of the lower angular resolution and the complexity/clumpiness of the molecular emission. Position-velocity (PV) diagrams are thus applied to investigate this shell structure, and are used to examine whether a shell appears as an expanding ring/bubble in the molecular line data. Linear PV diagrams of $^{13}$CO $J=2-1$ and C$^{18}$O $J=2-1$ are shown in Figure \ref{Fig2}, corresponding to the east-west cut (Cut I) and the north-south cut (Cut II) in Figure \ref{Fig1}, where the $^{13}$CO and C$^{18}$O counterparts to the proposed two MIR shells are revealed in the PV diagrams. In addition, clumps are located along the shells which suggests that they are formed within them. In order to estimate the mass and kinematic properties of the shell structure, we constructed a circular PV diagram which azimuthally averages spectra at a given radius from the center. The center was chosen to maximize the velocity difference between peaks in the C$^{18}$O $J=2-1$ spectra, and is close to the UC H{\sc ii} regions O and N. Moreover, the circular PV diagram (Fig. \ref{Fig2} c) shows a clear elliptical signature which implies an expanding motion centered on a common origin. Therefore, we conclude that the two MIR shells share a common expansion center with an average inner radius of $\approx 2.9$ pc ($10 \%$ uncertainty) and an expanding velocity of $\sim 5$ km s$^{-1}$. We can reject the possibility that the shells are actually molecular rings \citep{Beaumont2009} because two-component spectra are seen toward the center of the shells probing their front and back sides (Fig. \ref{Fig2}). In the LTE assumption, we can estimate the H$_2$ column density of the expanding shells from the $^{13}$CO/C$^{18}$O $J=2-1$ ratio, assuming their optical depth ratio is the same as their abundance ratio. We adopted $^{12}$C/$^{13}$C of 46 and $^{16}$O/$^{18}$O of 510 \citep{Wilson1994,Langer1990} for W49A \citep[8.1 kpc to the Galactic center,][]{Gwinn1992}. The excitation temperature of the averaged shell is estimated to be about 25 K. The average H$_2$ column density of the shell is $\approx 8.5\times10^{22}$ cm$^{-2}$, excluding the contribution from source G (i.e., consider the shell density within a radius of 40\arcsec\ in Fig. \ref{Fig2} d). Assuming a $10\%$ uncertainty, we calculate an H$_2$ density, $n$(H$_2$), of $\approx 6.7\pm2.1\times10^{3}$ cm$^{-3}$ in a spherical structure with a filling factor of 1 and a CO abundance of $2\times10^{-4}$ \citep[e.g.,][]{Lacy1994}. The shell mass ($M_{\rm{sh}}$) can be estimated as $1.9\pm0.7 \times10^{4}$ M$_{\sun}$. Then the kinetic energy ($\frac{1}{2}M_{\rm{sh}}V^2_{\rm{sh}}$) and the momentum ($M_{\rm{sh}}V_{\rm{sh}}$) can be computed to be $4.6\pm3.0 \times10^{48}$ erg and $9.3\pm4.0 \times10^{4}$ M$_{\sun}$ km s$^{-1}$, respectively, assuming a $20\%$ uncertainty in the expansion speed measurement. 


   \begin{figure*}
   \centering
   \includegraphics[angle=-90,width=\textwidth]{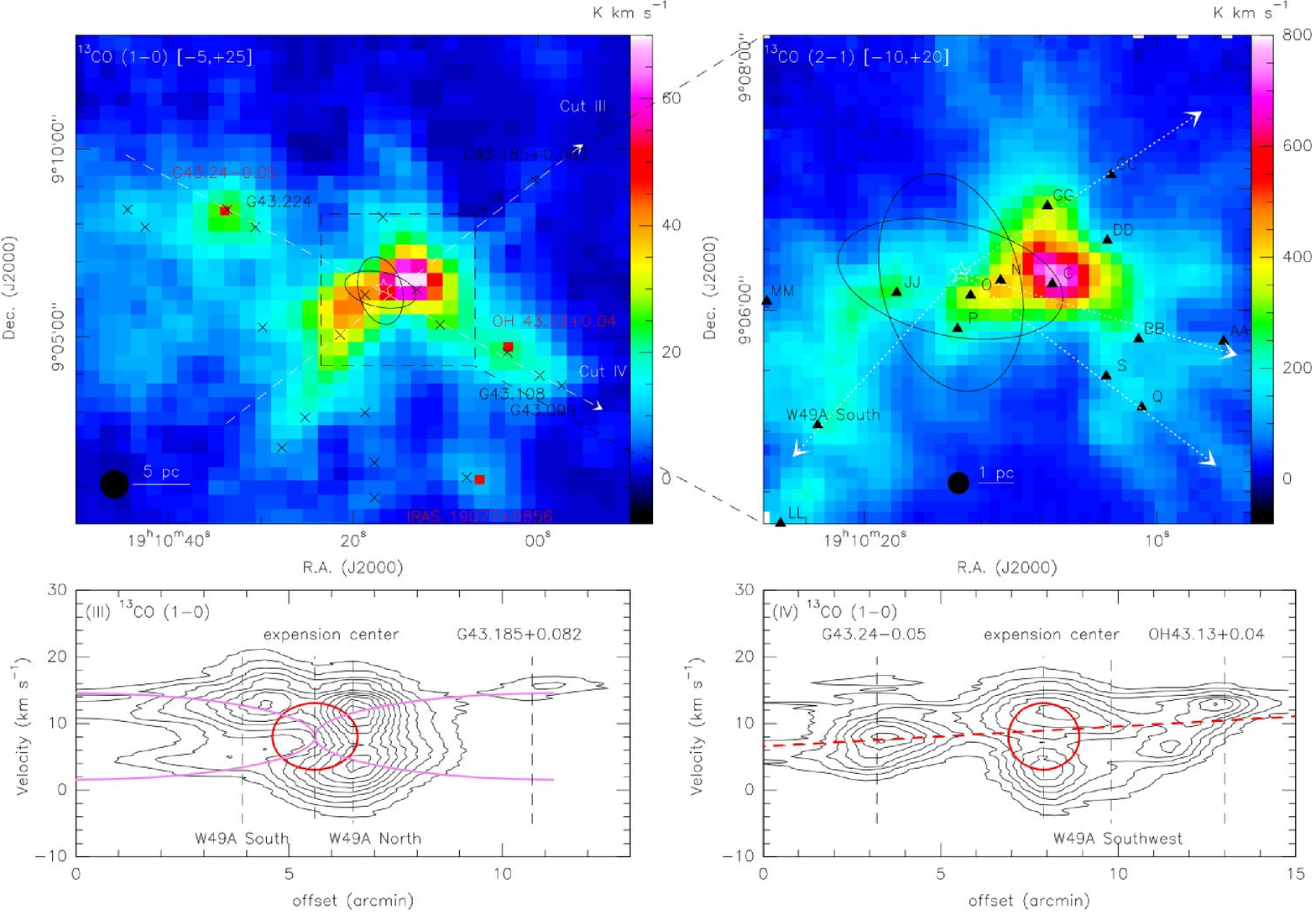}
   \caption{Upper panels: The left panel shows the large-scale GRS $^{13}$CO $J=1-0$ emission of the W49 complex, and the right panel shows the $^{13}$CO $J=2-1$ emission from IRAM 30m in the local area of W49A. The MIR shell structure is indicated by black ellipses. The white dashed lines represent two linear PV cuts shown in the lower panels. The white dotted lines represent the suggested alignments or ejections of molecular gas. Black crosses represent the 850 $\mu$m dust continuum peaks from \citet{Matthews2009}. Black triangles denote the UC H{\sc ii} regions from \citet{DePree1997}. (III)-(IV) The NW-SE and NE-SW PV diagrams corresponding to Cut III and Cut IV in the upper-left panel. The shell with a radius size of 3.3 pc and an expansion speed of 5 km s$^{-1}$ is indicated in red ellipses. The violet curves in a double-cone shape represent the gas ejections in the NW-SE direction, and the red dashed line represents the NE-SW direction of the bipolar gas ejections. The contours denote the temperatures from 0.4 to 4.8 K in steps of 0.4 K.}
              \label{Fig3}
    \end{figure*}

\section{Discussion}

\subsection[]{Driving mechanisms for the expanding shells}

The reason for the elliptical shape of the two shells is not clear, but is probably due to different expansion velocities in the two axes, or due to multiple expanding events. In the interstellar bubble model of \citet{Weaver1977}, where the spherical expanding shells are driven by winds, the age of an averaged expanding shell in W49A can be estimated as $3.4\pm0.4 \times10^{5}$ yr from the shell radius and expansion speed ($t=\frac{3}{5}\frac{R_{\rm{sh}}}{V_{\rm{sh}}}$). Here, the radius of the shell depends on the mass-loss rate $\dot{M}$ of the driving source \citep{Weaver1977}
\begin{eqnarray}
R_{\rm{sh}}(t)=28.1\left(\frac{\dot{E}}{10^{36}\ \rm{erg\ s^{-1}}}\right)^{1/5}\ \left(\frac{\mu\ n_{0}}{1\ \rm{cm^{-3}}}\right)^{-1/5}\ \left(\frac{t}{10^6\ \rm{yr}}\right)^{3/5}\ \rm{pc},
\end{eqnarray}
where $\dot{E}$ is the mechanical luminosity ($\dot{E}=\frac{1}{2}\dot{M}_wV_w^2$), and $n_{0}$ is the ambient H density. We derived an average H$_2$ density of $\sim 2\times10^{3}$ cm$^{-3}$ in W49A by using the $^{13}$CO/C$^{18}$O $J=2-1$ ratio mentioned above, and we adopted an ambient density $n_{0}$ of $4\times10^{3}$ cm$^{-3}$ and a stellar wind velocity $V_w$ of 2000 km s$^{-1}$. Hence, a constant mass-loss rate of $\sim 1.2\times10^{-6}$ M$_{\sun}$ yr$^{-1}$ is needed to sustain a wind-driven bubble in a size of $\sim2.9$ pc, which corresponds to the mass loss rate of one O-type star. In addition, about $20\%$ of the wind energy is channeled into the kinetic energy of the swept-up shell \citep{Weaver1977}, which means that a total energy of $\approx 3.4\pm\ 2.2\times10^{48}$ erg is provided by winds during the expansion. This is consistent with the average shell kinetic energy of $4.6\pm3.0 \times10^{48}$ erg. However, given the elliptical shape of the two shells, their expansion likely differs from that of the interstellar bubble model discussed above. Therefore, a reasonable upper-limit of the shell age ($t\leq\frac{R_{\rm{sh}}}{V_{\rm{sh}}}$) is estimated to be $5.7\pm1.3 \times10^{5}$ yr, assuming either a monotonically decreasing expansion velocity or a constant velocity. We conclude that the ages of these expanding shells are likely about $3$--$7\times10^5$ yr.


On the other hand, a high radiation pressure in H{\sc ii} regions can also provide enough energy to drive an expanding shell \citep[e.g.,][]{Krumholz2009}. \citet{Homeier2005} estimate a cluster mass of $\sim 1\times10^{4}$ M$_{\sun}$ around source O and source N by summing 54 stars with masses greater than 20 M$_{\sun}$ within 2.5 pc, which means that this cluster probably hosts $\sim30$--$50$ O6 stars with an ionization luminosity Q of $\sim 1\times10^{51}$ s$^{-1}$ \citep{Vacca1996}. This corresponds to a luminosity $L$ of $8\times10^{40}$ erg s$^{-1}$. The luminosity ($L=P_{\rm{sh}}c/t$) needed for energy-driven shells can be estimated to be $\sim 8\times10^{39}$ erg s$^{-1}$, assuming no photon leaking. Therefore, no matter whether the expanding shells are driven by winds or by radiation pressure, just a few massive stars can account for the driving energy rather than the whole cluster. A recent model of W49A by \citet{Murray2010} demonstrated that the dominating radiation pressure from a new born cluster forms a bubble which eventually disrupts the whole cloud. In their one-dimensional model, the outward force is dominated by the force from protostellar jets during the first $3\times10^5$ yr, and radiation pressure dominates until the first supernova explodes. However, in the early phases, the expanding shell is driven by protostellar jets in their model, and it is not clear how those jets can drive the shells in W49A. Furthermore, leaking photons may reduce the radiation force on the shell. It is possible that many of the embedded sources seen by \citet{Homeier2005} are stars formed after shell fragmentation. We suspect in fact that these two shells originate from a common center, which is close to the embedded cluster center indicated by \citet{Homeier2005}, and trigger star formation later when expanding.

\subsection[]{Fragmentation of the shells}

According to \citet{Whitworth1994}, an expanding shell begins to fragment at $t_{\rm{fragment}}\sim(G\rho_{o}\mathcal{M})^{-1/2}$, depending on the ambient density $\rho_0$ and the shell Mach number $\mathcal{M}$. The time scale for fragmentation, $t_{\rm{fragment}}$, is shorter than the time scale needed for the shell to become dominated by self-gravity, $t_{\rm{gravity}}\sim(G\rho_{\rm{sh}})^{-1/2}$. The value of $t_{\rm{fragment}}$ is estimated to be $\approx 3.6\times10^{5}$ yr, with a sound speed $C_{\rm{H_2}}$ of 0.35 km s$^{-1}$ and a Mach number of about 14 inside the shell. The time scale for the shells in W49A to fragment into clumps and further collapse to form new stars is about a few times $10^{5}$ yr, which roughly corresponds to the time scale of massive star formation \citep{Zinnecker2007}. The clumps formed after fragmentation are well separated as shown in Figure \ref{Fig2}, and the two velocity components at $\sim 4$ km s$^{-1}$ and $\sim 12$ km s$^{-1}$ toward the region around source G are perhaps two clumps along the line of sight. 

\subsection[]{Large-scale gas ejections}

A recent paper by \citet{Matthews2009} reveals the distribution of dust clumps in the large-scale environment of the W49 complex, including one UC H{\sc ii} region G43.24--0.05 \citep{Kurtz1994} and one hydroxyl maser source OH43.13+0.04 \citep{Becker1992}. Many dust clumps are aligned toward the region near source O and source N, which are close to the center of our expanding shells (Fig. \ref{Fig3}). Furthermore, the PV diagrams derived using the Galactic Ring Survey \citep[GRS;][]{Jackson2006} $^{13}$CO $J=1-0$ emission show an ``outflow structure'' which may be related to the expanding shells. For example, the PV diagram of Cut IV shown in Figure \ref{Fig3} indicates a bipolar outflow between G43.24--0.05 and OH 43.13+0.04 with the center of the expanding shell structure located roughly symmetrically between the two clumps seen in mm emission. Other evidence is the detection of UC H{\sc ii} regions (Fig. \ref{Fig3}) toward the shell expansion center as well as the fact that the filaments seen in $^{13}$CO $J=2-1$ have a radial geometry relative to the shell expansion center (although those filaments may be also related to W49A North or source G). Still further evidence of gas ejections comes from a large-scale gas distribution in the NW-SE direction suggesting that widely-distributed molecular gas and UC H{\sc ii} regions are present together with the double-cone structure shown in Figure \ref{Fig3} (III).


Without taking the inclination into account, we can estimate the projected kinetic energy (as a lower-limit) of the NE-SW gas ejections by assuming that they occurred at the same time when the shells started expanding (i.e., $5.7\times10^5$ yr as an upper-limit) and by adopting the clump mass estimates of \citet{Matthews2009}. In the northeastern ejection, the clumps G43.24--0.05 and G43.224--0.038 have a total projected kinetic energy of $ 1.3\pm0.6 \times10^{50}$ erg; in the southwestern ejection, W49A Southwest, OH43.13+0.04, submillimeter sources G43.108+0.044 and G43.099--0.050 have a total projected kinetic energy of $1.1\pm0.3 \times10^{50}$ erg, assuming a $20 \%$ uncertainty in the mass measurements. The projected kinetic energy estimates are summarized in Table 1. The projected kinetic energy for both directions is roughly equal, and we infer a total projected kinetic energy of $2.4\pm0.7 \times10^{50}$ erg. This result depends on our assumption about the age of the outflows. If the gas ejection time scale were seven times longer than the age of expanding shells, the kinetic energy for the shells and the outflows would be similar. In either case, the total energy released by shell expansion and gas ejection has a magnitude of $10^{50}$ erg. Since we only estimated the kinetic energy for the NE-SW gas ejection, the total kinetic energy of the gas ejection event in the W49 complex may be larger. It is worth noting that G43.24--0.05 is located at a similar distance \citep[11.7$^{+0.8}_{-0.7}$ kpc;][]{Watson2003} as W49A, and probably hosts a newly born star producing the UC H{\sc ii} region. In addition, the nearby source G43.224-0.038 is likely related to G43.24--0.05. The nature of the OH maser source OH43.13+0.04 is not clear, since it is not associated with any of the Infrared Astronomical Satellite (IRAS) sources \citep{Becker1992}. Additionally, the nature of the submm sources G43.108+0.044 and G43.009+0.050 are also unknown.


\begin{table}
\caption{The projected kinetic energy estimates of gas ejections}             
\label{table:1}      
\centering                          
\begin{tabular}{l c c c c}        
\hline\hline                 
Source & $D^{\rm a}$  & $M^{\rm b}$          & $V^{\rm c}$           & $E_{\rm kin}$ \\    
       & (pc)         & (M$_{\sun}$)         & (km s$^{-1}$)         & ($\times10^{49}$ erg) \\
\hline                        
G43.224--0.038    &  12.6  &  $2.6\times10^3$  &  $21.7\pm4.9$  &  $1.2\pm0.6 $ \\     
G43.24--0.05      &  15.6  &  $1.6\times10^4$  &  $26.8\pm6.0$  & $11.4\pm5.6 $ \\     

\hline 
W49A Southwest   &   6.2  &  $3.5\times10^4$  &  $10.7\pm2.4$  & $4.0\pm2.0 $ \\    
OH43.13+0.04     &  12.2  &  $8.0\times10^3$  &  $21.0\pm4.7$  & $3.5\pm1.7 $ \\    
G43.108+0.044    &  15.9  &  $1.9\times10^3$  &  $27.4\pm6.1$  & $1.4\pm0.7 $ \\    
G43.009+0.050    &  18.0  &  $2.0\times10^3$  &  $31.0\pm6.9$  & $1.9\pm0.9 $ \\    

\hline                                   
\end{tabular}

\begin{list}{}{}
\item[$^{\mathrm{a}}$] The distance to the shell expansion center without a projection correction.
\item[$^{\mathrm{b}}$] The gas mass of clumps adopted from \citet{Matthews2009}.
\item[$^{\mathrm{c}}$] The ejection velocity of clumps assuming a constant ejection velocity in $5.7\times10^5$ yr.

\end{list}

\end{table}

The magnetic field measurement by \citet{Brogan2001} shows a gradient in the line-of-sight component of the magnetic field, $B_{\rm{los}}$, in the NE-SW direction crossing the expansion center, where a very weak $B_{\rm{los}}$ is derived from the 7 km s$^{-1}$ H{\sc i} component. For a radially increasing magnetic field, a weak $B_{\rm{los}}$ is expected in the shell expansion center ($V_{\rm{LSR}}\sim7$--8 km s$^{-1}$). The fact that $B_{\rm{los}}$ increases in the northeastern and southwestern directions may be related to the bipolar outflow in the NE-SW direction. In addition, another strong $B_{\rm{los}}$ peak in the northwestern direction at 4 km s$^{-1}$ could be related to the NW-SE gas ejection with a double-cone shape. It is not clear what kind of event can cause large-scale gas ejections with a kinetic energy of a few times $10^{50}$ erg. A supernova explosion is unlikely since such an explosion would destroy the whole cloud with a kinetic energy of $\sim10^{51}$ erg. A stellar merger of two 30 M$_{\sun}$ stars can release a kinetic energy as high as $\sim 2\times10^{50}$ erg \citep{Bally2005}. However, there is no evidence for such an event, and it may be impossible to find conclusive evidence. Thus, understanding the driving source requires examining the region around the shell expansion center in detail.

\section{Conclusions}

The morphology of the molecular line emission and the associated velocity signatures are consistent with the shell structure seen in the GLIMPSE images. The expanding shells in W49A are suggestive of triggered massive star formation in just $\sim 10^5$ yr. There is evidence for dense clumps coincident with the shells, and the two velocity components toward the region around source G may be two distinct clumps along the line of sight. The expanding shells are likely to have a common origin close to source O with an expansion speed of $\sim5$ km s$^{-1}$ and an age of 3--$7\times10^{5}$ years. However, it is not clear what is the driving mechanism of the shell expansion. The average mass of the expanding shells is $\sim2\times10^{4}$ M$_{\sun}$ with a kinetic energy of $\sim10^{49}$ erg, which is probably powered by a few massive stars in the center of the cluster, and the expanding shells are either driven by strong winds or by radiation pressure. In addition, evidence of the gas ejections is present in the large-scale GRS $^{13}$CO $J=1-0$ and dust continuum data. Both ejections appear to have the same origin as the expanding shells with a total energy of a few times $10^{50}$ erg. The properties of the central source driving the expanding shells and gas ejections need further investigation.

\begin{acknowledgements}

We thank the IRAM 30m staff for the support during the observations. This work was supported by the International Max Planck Research School (IMPRS) for Astronomy and Astrophysics at the Universities of Bonn and Cologne. We have made use of the NASA/IPAC Infrared Science Archive to obtain data from the Spitzer-GLIMPSE. This publication makes use of molecular line data from the Boston University-FCRAO Galactic Ring Survey (GRS). The GRS is a joint project of Boston University and Five College Radio Astronomy Observatory, funded by the National Science Foundation under grants AST-9800334, AST-0098562, AST-0100793, AST-0228993, \& AST-0507657.
\end{acknowledgements}

\end{document}